\documentclass[]{aa}

\usepackage{verbatim} 
\usepackage{graphicx}
\usepackage{txfonts}
\usepackage{natbib}        
\bibpunct{(}{)}{;}{a}{}{,} 

\newcommand{\baseA}{\ensuremath{\vec{B}_{12}}}
\newcommand{\baseB}{\ensuremath{\vec{B}_{23}}}
\newcommand{\baseC}{\ensuremath{\vec{B}_{31}}}
\newcommand{\closure}{\ensuremath{\phi}}
\newcommand{\magnif}{\ensuremath{m}}
\newcommand{\position}{\ensuremath{\vec\Delta}}
\newcommand{\ratio}{\ensuremath{\rho}}
\newcommand{\baseAn}{\ensuremath{\alpha_{12}}}
\newcommand{\baseBn}{\ensuremath{\alpha_{23}}}
\newcommand{\baseCn}{\ensuremath{\alpha_{31}}}

\newcommand{\wave}{\ensuremath{\lambda{}}}

\begin{document}

\title{On the sensitivity of closure phases to faint companions in optical long baseline interferometry}
\titlerunning{Sensitivity of closure phases to faint companions}

\author{J.-B.~Le~Bouquin\inst{1} \and O.~Absil\inst{2}\fnmsep\thanks{Postdoctoral Researcher F.R.S.-FNRS (Belgium)}}

\institute{
  UJF-Grenoble 1 / CNRS-INSU, Institut de Plan{\'e}tologie et d'Astrophysique de Grenoble (IPAG) UMR 5274, Grenoble, France
  \and
  D\'epartement d'Astrophysique, G\'eophysique et Oc\'eanographie, Universit\'e de Li\`ege, 17 All\'ee du Six Ao\^ut, 4000 Li\`ege, Belgium
}
\offprints{J.B.~Le~Bouquin\\
  \email{jean-baptiste.lebouquin@obs.ujf-grenoble.fr}}
  
\date{Received 16/08/2011; Accepted 15/03/2012}

\abstract
{}
{We explore the sensitivity and completeness of long baseline interferometric observations for detecting unknown, faint companions around bright unresolved stars.}
{We derive a linear expression for the closure phase signature of a faint companion in the high contrast regime ($\leq0.1$), and provide a quantitative estimation of the detection efficiency for the currently offered four-telescope configurations at the Very Large Telescope Interferometer. The results are compared to the performances provided by linear and Y-shaped interferometric configurations in order to identify the ideal array.}
{We find that all configurations have a similar efficiency in discovering companions wider than 10\,mas. Assuming a closure phase accuracy of $0.25\deg$, that is typical of state-of-the-art instruments, we predict a median dynamic range of up to six magnitudes when stacking observations obtained at five different hour angles.}
{Surveying bright stars to search for faint companions can be considered as an ideal filler programme for modern interferometric facilities because that places few constraints on the choice of the interferometric configuration.}


\maketitle


\section{Introduction}
The most successful method for detecting faint companions around nearby stars is undoubtedly the radial velocity (RV) technique \citep{Perryman00}, with more than 500 extrasolar planets detected up to date. Nevertheless, RV has three major drawbacks. First, it cannot be applied to all kind of stars: pulsating stars, as well as young, rapidly rotating and/or active stars have an intrinsic radial velocity jitter that generally precludes planet-search programmes \citep{Udry07}. The discovery of extrasolar planets orbiting high-mass main-sequence stars (O, B, and A spectral types) and young solar-type stars (T Tauri stars) therefore remains rare in the literature. Second, RV can only access companions with periods shorter than a few tens of years, owing to the limited time span of the observations obtained with high-precision spectrographs. Third, RV is inherently insensitive to the orbital inclination $i$. This implies that additional astrometric observations are mandatory to unveil the real companion mass from the measured quantity $M_p\sin{i}$. For these reasons, the development of direct imaging techniques for low-mass companions is currently one of the highest priorities in instrumental astronomy \citep{Oppenheimer09,Absil10b}.

Direct imaging techniques should be tailored to the typical angular separations for which low-mass companions are to be found. For instance, the search for young planets in nearby star-forming regions ($\geq 100$\,pc) should be optimised for typical linear separations from 0.1 to a few 10\,AU for Sun-like stars, where planets are supposed to be forming and migrating in their early history \citep{Bodenheimer02}. This translates into typical angular separations of 1--100\,milli-arcsecs (mas). In a more general context, more than 90\% of the sub-stellar companions detected to date by all techniques have expected apparent separations smaller than 100\,mas \citep{Schneider11}. This domain is hardly accessible to single-pupil imaging techniques, even with adaptive-optics (AO) assisted facilities \citep{Burrows05}. Nowadays, with 10-m class telescopes, only interferometric aperture-masking observations routinely break the 100\,mas limit. However, their search space is still restricted to angular separations larger than about 40\,mas for dynamic ranges of the order of 500:1 \citep{Kraus08,Lacour:2011}.

Thanks to its higher angular resolution, optical long baseline interferometry is the ideal tool for exploring separations in the range 1--50\,mas. The simplest strategy for faint companion detection with optical/infrared interferometry is to obtain closure phase measurements that are directly sensitive to the asymmetry in the brightness distribution of the target source and hence to possible off-axis companions. However, owing to the sparse structure of the point spread function associated with the diluted aperture of an interferometer, the depth to which a companion can be detected strongly depends on the relative orientation of the companion and the interferometric baselines. Consequently, the sensitivity limit should be defined as a two-dimensional map and/or for various completeness levels.

The goal of this paper is to provide quantitative estimations of the detection efficiency versus the companion contrast and separation considering realistic observations. In Sect.~\ref{sec:theory}, we derive a linear expression for the closure phase signature of a faint companion. We also define a simple, general expression for the sensitivity limit in terms of companion contrast, which we use in Sect.~\ref{sec:array} to compute the efficiency of various interferometric configurations. We compare the efficiency of the four configurations currently offered at the Very Large Telescope Interferometer. To add generality, we also discuss the efficiency of three ``standard'' interferometric arrays (two linear and one Y-shaped). In Sect.~\ref{sec:accuracy}, we discuss the typical achievable dynamic range for the closure phase accuracy provided by existing instruments.


\section{Detecting faint companions with closure phase}
\label{sec:theory}

Optical long baseline interferometers provide several observables, which are all sensitive to the presence of a faint companion: visibilities, differential phases, and closure phases. The latter has the main advantage of being, to first order, uncorrupted by telescope-specific phase errors, including pointing errors, atmospheric piston, and longitudinal dispersion due to air and water vapour. The noise floor is only limited by the accuracy to which the instrumental systematics can be mastered. As a consequence, the high-precision closure phase is the favoured observational strategy to directly detect faint companions with modern interferometers \citep{Vannier06,Zhao08,Renard08}.

\subsection{Closure-phase signal of a faint companion} \label{sec:signal}

Measuring a closure phase requires the use of an interferometric array composed of (at least) three telescopes. Each pair of telescopes defines a geometrical vector referred to as the interferometric baseline. The closure phase is then defined as the sum of the phases measured on the three baselines, or equivalently, as the argument of the bispectrum, formed through the triple product of the measured complex visibilities around the triangle \citep{Monnier03}. Assuming that the stellar diameters are not resolved at the considered interferometric baseline lengths, the closure phase signature of a binary object can be expressed as a function of the three baseline vectors projected onto the sky \{\baseA,\baseB,\baseC\}, the wavelength of observation \wave{}, the binary flux ratio \ratio{} (referred to as the contrast in the following), and the apparent binary separation vector $\position$
\begin{equation}
\closure = \arg \left( \frac{(1+\ratio\,e^{i\baseAn}) (1+\ratio\,e^{i\baseBn}) (1+\ratio\,e^{i\baseCn})} {(1+\ratio)^3} \right) \, , \label{eq:1}
\end{equation}
where
\begin{equation}
\baseAn = 2\pi \frac{\baseA \cdot \position}{\wave} \; ; \;\;
\baseBn = 2\pi \frac{\baseB \cdot \position}{\wave} \; ; \;\;
\baseCn = 2\pi \frac{\baseC \cdot \position}{\wave} \; . \label{eq:1b}
\end{equation}
The geometrical terms \baseAn{}, \baseBn{}, and \baseCn{} describe the relative orientation of the companion compared to the spatial frequencies explored by the interferometric array. We note that $\baseCn=-(\baseAn+\baseBn)$ since the vectorial sum of the three interferometric baselines is zero by definition.

To model the signal of a high contrast binary, we can either use Eq.~\ref{eq:1} with small ($\ll$$1$) or large ($\gg$$1$) values of $\ratio$. In the first case, the interferometer points toward the primary, while in the second case it points toward the companion. Since the closure-phase is independent of the pointing position, these two approaches are formally identical. Assuming $\ratio\ll1$, we can develop and simplify Eq.~\ref{eq:1} to be
\begin{equation}
\closure \approx \ratio \, (\sin\baseAn + \sin\baseBn - \sin(\baseAn+\baseBn)) \, . \label{eq:2}
\end{equation}
The quantity $\magnif=\sin\baseAn + \sin\baseBn - \sin(\baseAn+\baseBn)$ is referred to as the magnification factor because a high value of $\magnif$ corresponds to a stronger companion signature in the closure phase signal.

The validity range of our approximate formula can be explored numerically by computing the difference between Eq.~\ref{eq:1} and~\ref{eq:2} for various values of \{\ratio,\baseAn,\baseBn\}. We found that Eq.~\ref{eq:2} stays within a factor $\lesssim 1.5$ of the value given by Eq.~\ref{eq:1} as long as $\ratio \leq 10^{-1}$, which we consider to be the validity range of our work. Therefore, the following results apply only to relatively faint companions, and cannot be immediately transposed to binaries with flux ratios close to unity.

Apart from the computational gain, the main advantage of using Eq.~\ref{eq:2} is to define a proportional relation between the closure phase and the companion contrast. This reduces the number of parameters to be explored and allows the results to be presented in a synthetic way. Additionally, we note that in Eq.~\ref{eq:2}, the magnification factor $m$ takes a maximum value of about 2.6\,rad ($\approx149\deg$) for purely geometrical reasons. This magnification is only achieved in the optimal geometrical case, where all interferometric baselines add together to amplify the companion signal. Therefore, as a quantitative example, we can already conclude that reaching a dynamic range of $10^{-3}$ with one single closure phase measurement requires a closure phase accuracy \emph{better} than $0.15\deg$. In practice, this accuracy is generally reached after appropriate uv-plane or temporal averaging of a series of individual closure phase measurements.

\subsection{The close-companion limit}
In the specific case where the binary is not resolved by any individual interferometric baselines (all $\alpha<1$), it is possible to simplify Eq.~\ref{eq:2} by expanding the sines for small values of $\alpha$
\begin{equation}
  \frac{\phi}{\ratio} \approx \frac{\baseAn^2\,\baseBn + \baseAn\,\baseBn^2}{2} \;+\; o(\alpha^5) \; .
\label{eq:close1}
\end{equation}
We can arbitrarily choose $\baseAn$ to be associated with the most resolving baseline and $\baseBn{}$ with the least resolving baseline in the direction of the considered binary. We introduce the $r$ ratio of the two geometrical terms: $\baseBn=-r\,\baseAn$ (the negative sign is introduced for the triangle to close with positive values of $r$). Rewriting Eq.~\ref{eq:close1}, we obtain
\begin{equation}
  \frac{\phi}{\ratio} \approx \frac{1}{2}\,\baseAn^3\,r\,(1-r)
  \label{eq:close2}
\end{equation}
This equation already shows two interesting aspects: (i) the closure phase signature of an unresolved companion is proportional to the baseline length at the third power, and (ii) the arrays that are redundant when projected onto the binary direction $(r=0.5)$ seem optimal for detecting a close companion using a given long baseline.

\subsection{Deriving sensitivity limits}
A typical interferometric observation is not composed of a single closure phase measurement. We now derive sensitivity limits in terms of companion contrast for a set of $n$ linearly independent\footnote{Using $m$ apertures, one can form $(m-1)(m-2)/2$ linearly independent closure phases. This is equivalent to holding one aperture fixed and forming all possible triangles with that aperture \citep{Monnier03}.} closure phase measurements $\closure_i$ of accuracy $\sigma_i$. We base our analysis on the $\chi^2$ of the data with respect to a model of an unresolved source with no companion, for which all closure phases are zero
\begin{equation}
\chi^2 = \sum_{i=1}^n \frac{\closure^2_i}{\sigma_i^2} \; .
\end{equation}
We introduce the simplification presented in Eq.~\ref{eq:2} and assume that all the measurements have similar accuracies (a valid assumption when observing an unresolved target) to obtain
\begin{equation}
\chi^2 = \frac{\rho^2}{\sigma^2} \; \sum_{i=1}^n m_i^2 \; .
\end{equation}
The probability that the data set is compatible with the single-star model is given by the complement of the cumulative probability distribution function (CDF) with $n$ degrees of freedom
\begin{equation}
P = 1-\mathrm{CDF}_n(\chi^2) \; .
\end{equation}
If $P$ is below a predefined threshold, the data set allows the model with no companion to be rejected. The threshold can generally be fixed at a 3-$\sigma{}$ level, i.e., at a probability of 0.27\%. The choice of the threshold actually depends on the context of the observations. If a large region of parameter space (or indeed number of targets) is searched for a companion, then a 5-$\sigma{}$ threshold may be more appropriate. A higher threshold may even be needed in the case of sparse data sets dominated by non-Gaussian systematic errors. Taking the 3-$\sigma{}$ level as an example, this threshold can be converted into a sensitivity limit in terms of companion contrast
\begin{equation}
  \rho = \sigma \sqrt{\frac{\mathrm{CDF}^{-1}_n(1-0.27\%)}{\sum_{i=1}^n m_i^2}} \; ,
\end{equation}
where $\mathrm{CDF}^{-1}_n(1-0.27\%)$ is simply the $\chi^2$ value for a 3-$\sigma{}$ detection with $n$ degrees of freedom. As a consequence of the approximation presented in Eq.~\ref{eq:2}, the sensitivity limit is directly proportional to the accuracy of the closure phase measurements.

Because of the sparse structure of the point spread function associated with the diluted aperture of an interferometer, the depth to which a companion can be detected strongly depends on the relative orientation between the companion and the interferometric baselines (information embedded in the magnification factors $m_i$). For some lucky separations, the greatest dynamic range is achieved, while in the worst cases even obvious binaries with equal brightnesses can be missed. Consequently, the sensitivity limit should be defined as a two-dimensional map or for various completeness levels on a given search region.

\subsection{Validity limit of this study}
In the case of wide companions, care should be taken regarding the chromaticity limit of our study. The results presented here are indeed formally valid only for monochromatic light. The main effect of wavelength smearing inside Eq.~\ref{eq:1} is to degrade the dynamic range. To avoid significant smearing, the spectral resolution should be higher than the $\alpha$ quantities defined in Eq.~\ref{eq:1b}. This translates into a spectral resolution $R>10$ for a $40\,$mas binary observed with a $100\,$m baseline at $1.7\,\mu$m (H band). Such a low spectral resolution is available in most modern interferometric beam combiners. In the case of spatially filtered beam combiners, a similar effect may occur because of baseline smearing, in the case where the telescope size cannot be neglected in Eq.~\ref{eq:1}. For the $1.8$-m Auxiliary Telescopes, this corresponds to angular separations of about $175\,$mas for H-band observations. For $10$-m class telescopes, this corresponds to angular separations of about $45\,$mas. In practice, these limitations are not very relevant to our study because AO-assisted spare aperture-masking imaging on 10-m class telescopes becomes more efficient than long-baseline interferometry for separations larger than about $40\,$mas \citep[see e.g.][]{Kraus08,Lacour:2011}.

Finally, Eq.~\ref{eq:1} assumes that the angular diameters of the binary components are unresolved by the interferometric baselines. Resolving the diameter of the faint component indeed appears unrealistic, although maybe not for the central star, especially in the case of bright late-type giants or very nearby stars. In this latter situation, Eq.~\ref{eq:1} underestimates the closure phase signal, which peaks for a fully resolved primary star. This effect, referred to as \emph{closure phase nulling}, can lead to larger magnification factors than the limit $m<149\deg$ presented in Sect.~\ref{sec:signal}. This specific observing technique is discussed in detail in \citet{Chelli:2009}, while on-sky applications can be found in \citet{Monnier:2006}, \citet{Lacour08}, \citet{Zhao08}, and \citet{Duvert:2010}.


\section{Optical interferometric array}
\label{sec:array}

We use the formalism introduced previously to compute and compare the capabilities of various four-telescope interferometric configurations.

\subsection{Performances of VLTI configurations}

\begin{figure}
    \centering 
    \includegraphics[scale=0.58]{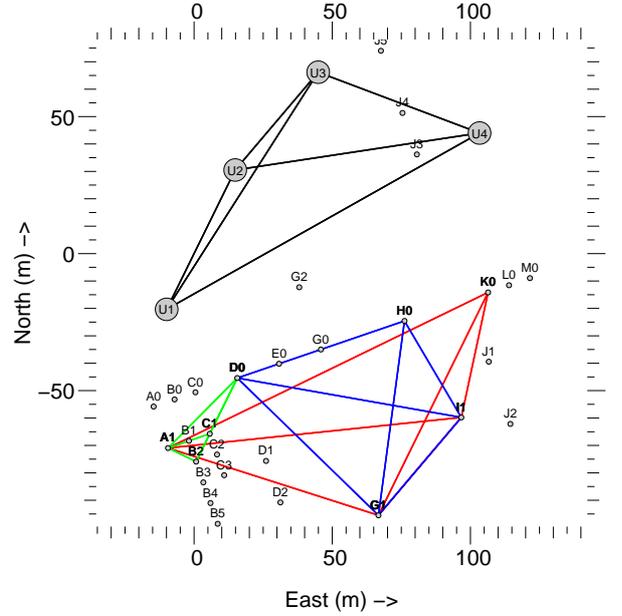}
    \caption{Interferometric configuration offered at VLTI. The configurations using the four relocatable Auxiliary Telescopes are represented by colours. The configuration using the four fixed Unit Telescopes is represented in black.}
    \label{fig:visa}
\end{figure}

\begin{figure*}
  \centering 
  \includegraphics[scale=0.52]{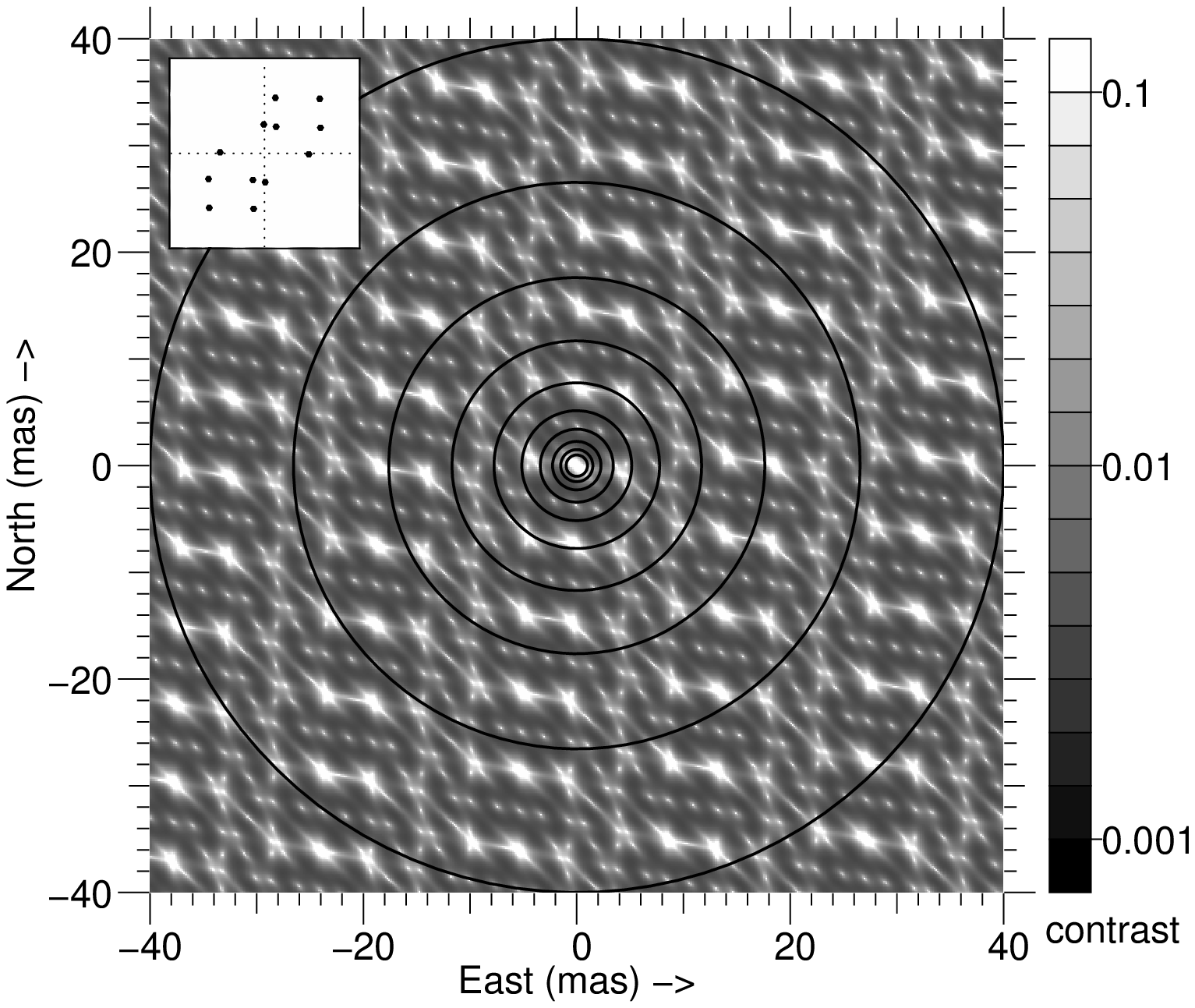}\hspace{0.7cm}
  \includegraphics[scale=0.49]{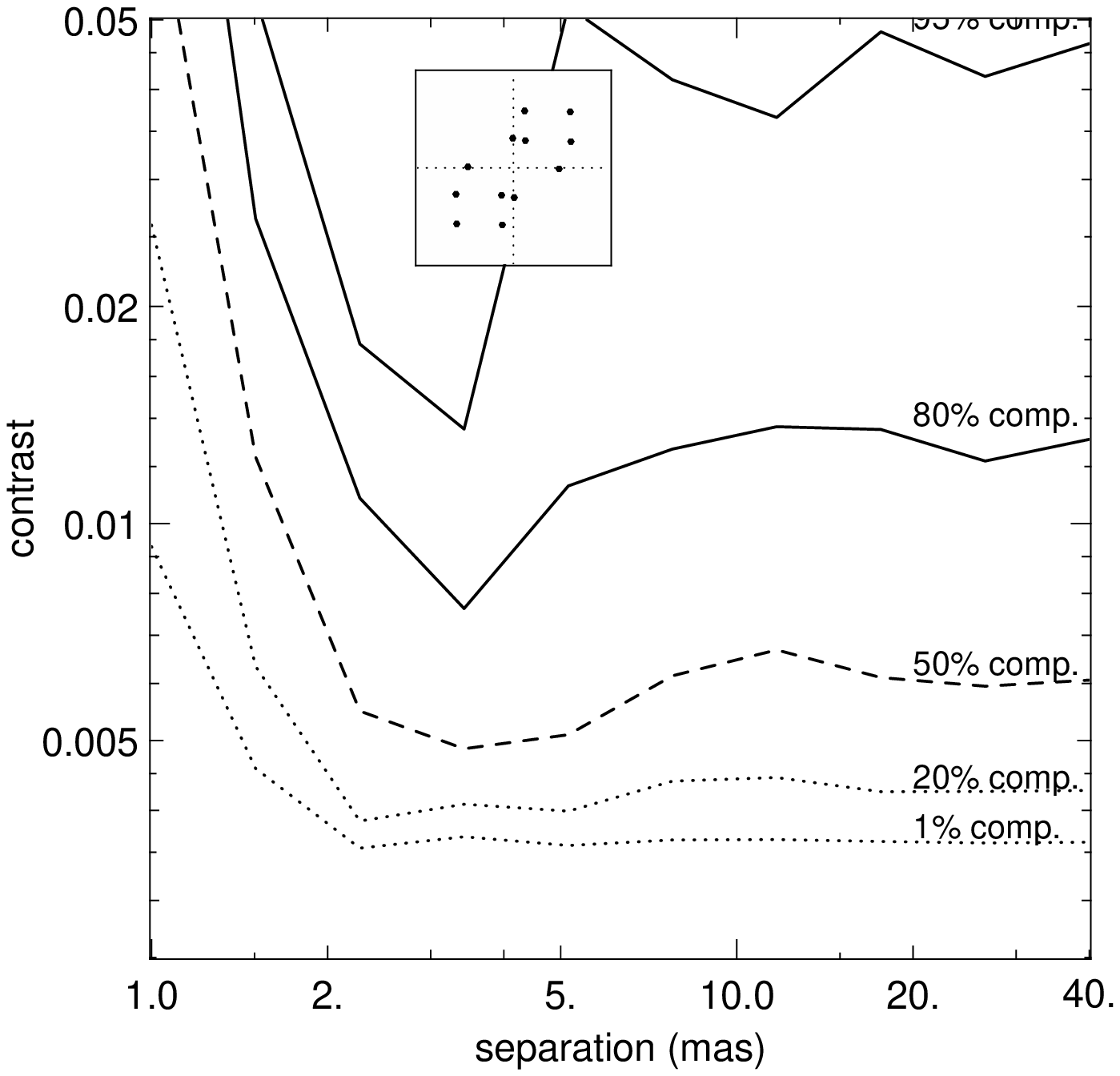}
  \includegraphics[scale=0.52]{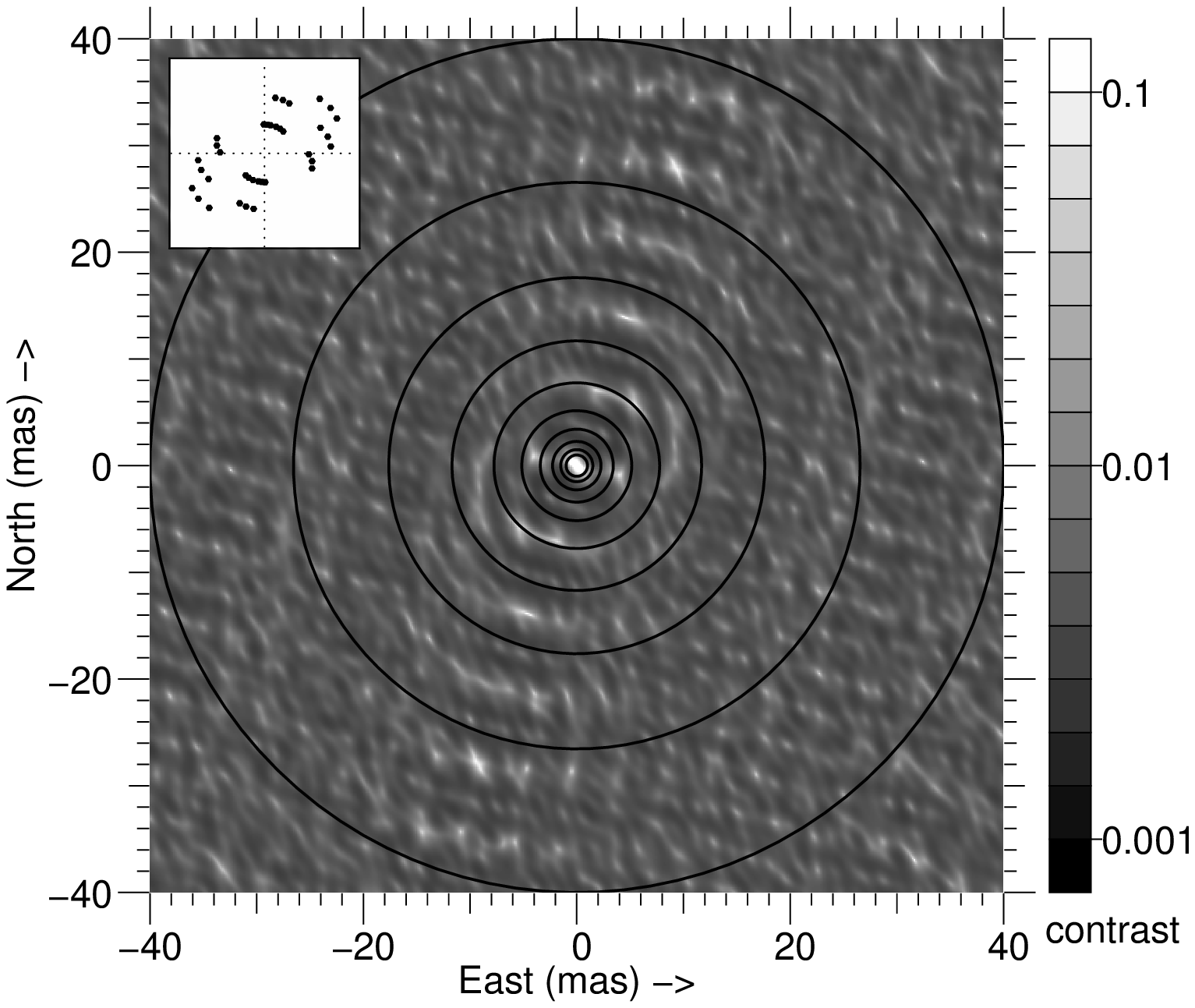}\hspace{0.7cm}
  \includegraphics[scale=0.49]{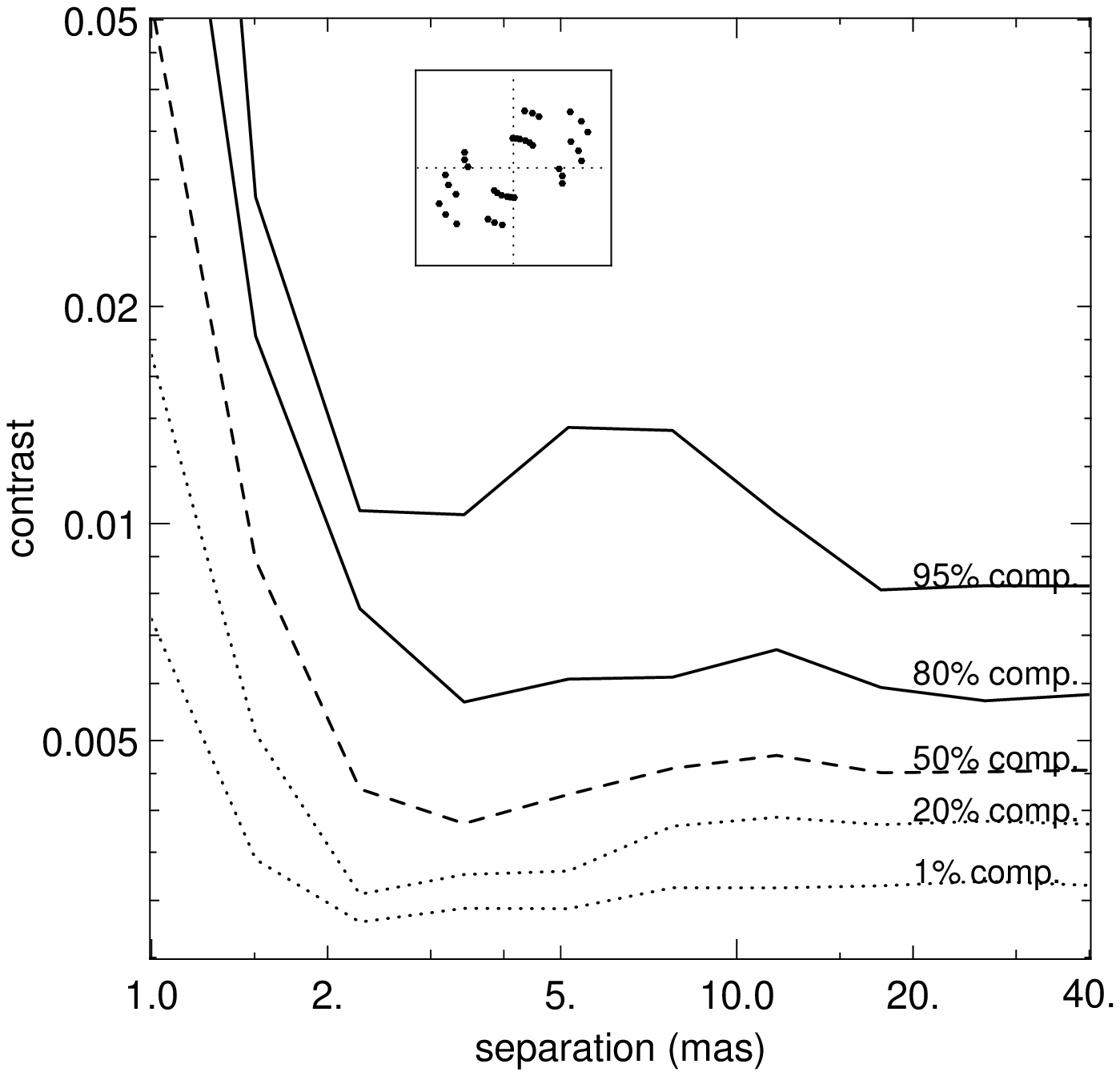}
  \includegraphics[scale=0.52]{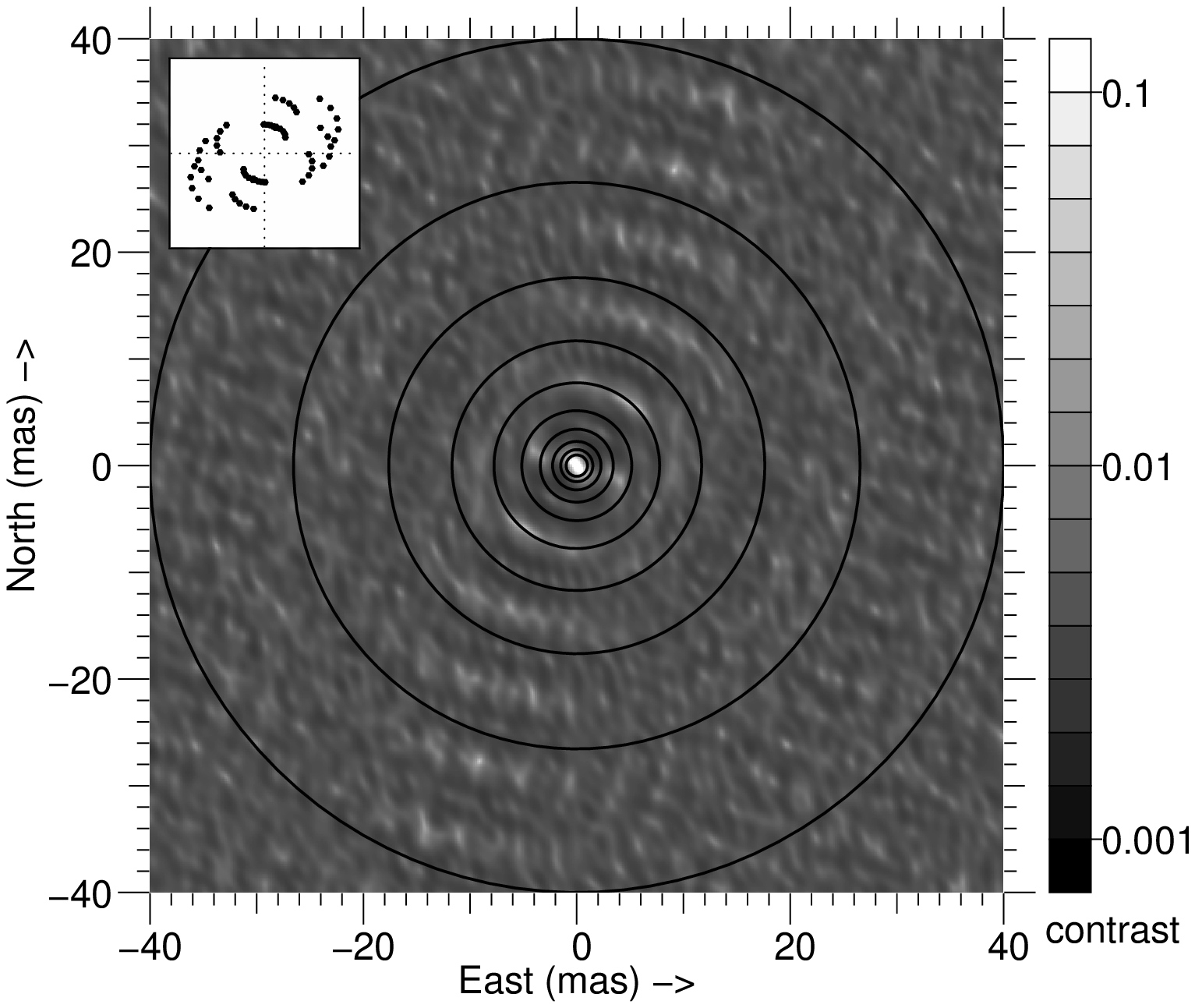}\hspace{0.7cm}
  \includegraphics[scale=0.49]{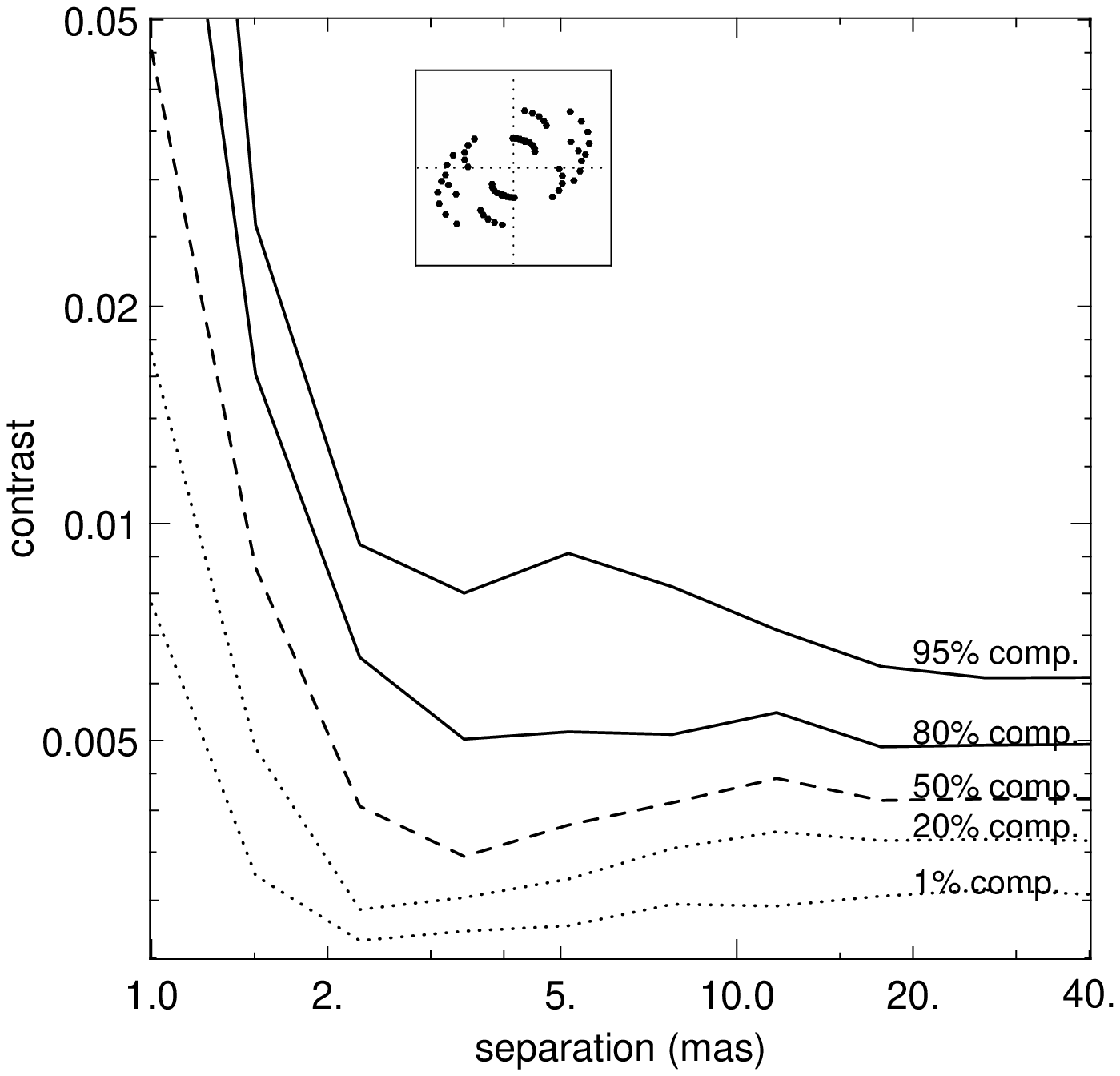}
  \caption{{\it Left:} Map of the 3-$\sigma{}$ sensitivity with the A1-K0-G1-I1 configuration from VLTI.  Radial zones are the bins in separation used for the plots in the right panel. {\it Right:}  Sensitivity as a function of angular distance, for various completeness levels. {\it From top to botom:}  Simulations for a single snapshot pointing (top), and for three pointing (middle), and for five pointing (bottom). The contrast axes can be scaled for any accuracy on the closure phase (here $\sigma=0.25\deg$).}
  \label{fig:det_map}
\end{figure*}

\begin{figure*}
    \centering 
    \includegraphics[scale=0.55]{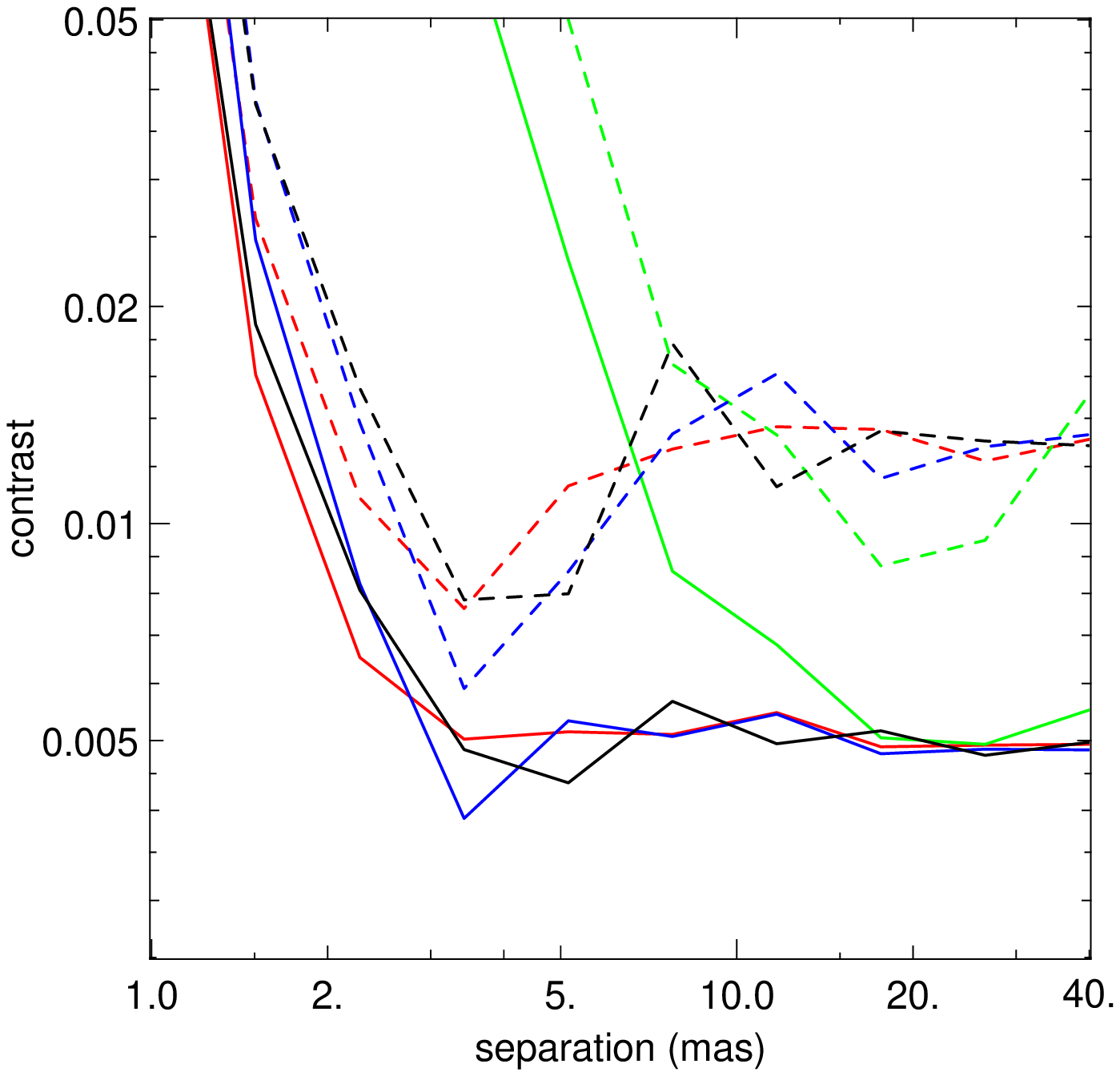}\hspace{2cm}
    \includegraphics[scale=0.55]{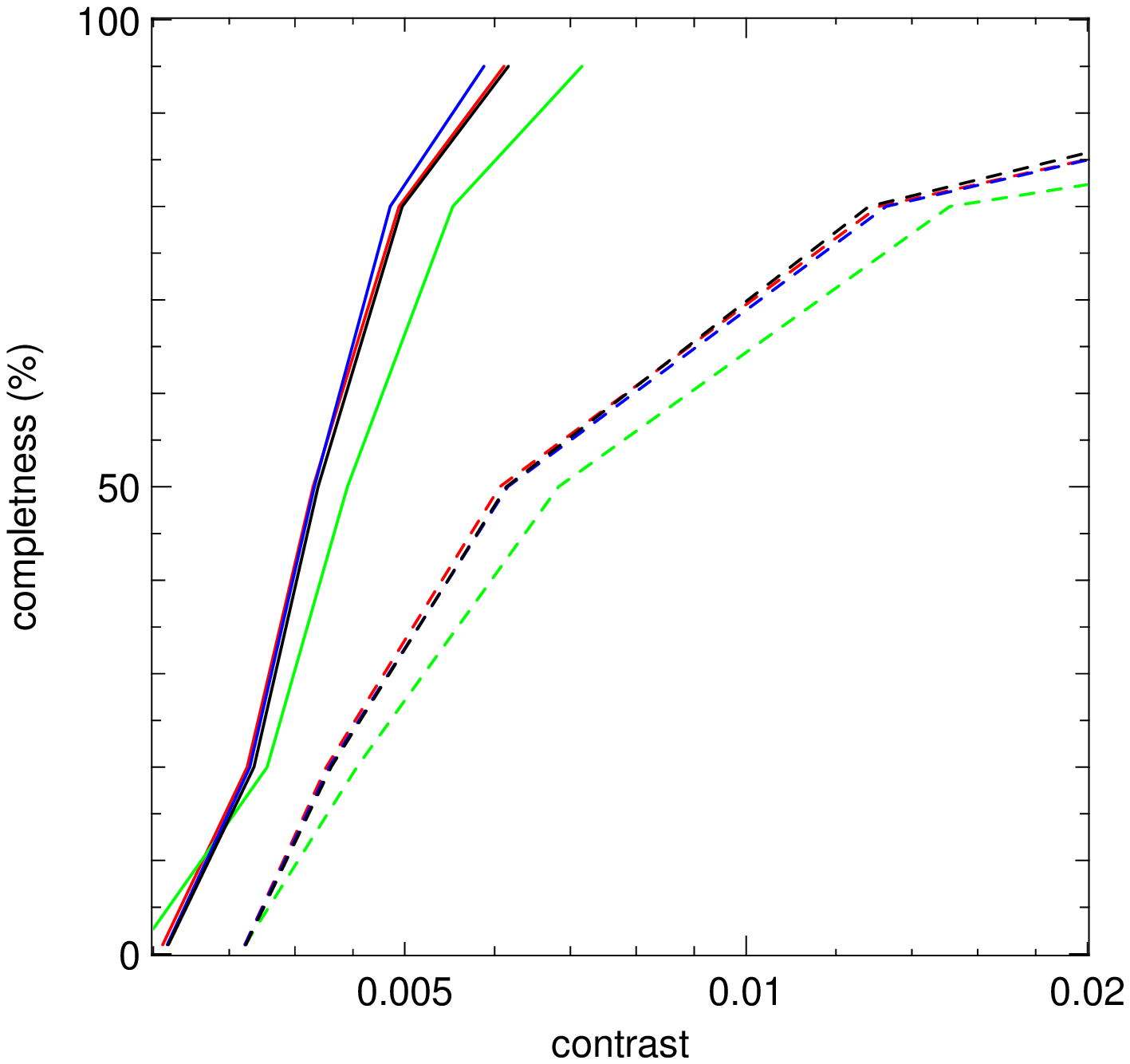}
    \caption{{\it Left:} 3-$\sigma{}$ sensitivity versus separation for a completeness level of 80\%. {\it Right:} Completeness in the separation range $6-40\,$mas versus contrast. Simulations are for a single snapshot pointing (dashed lines) and for five pointings separated by one hour (solid lines). Colours are for the four configurations of VLTI displayed in Fig.~\ref{fig:visa}. The contrast axes can be scaled for any given accuracy of the closure phase measurements (here $\sigma=0.25\deg$).}
    \label{fig:cont_sep}
    \label{fig:comp_cont}
\end{figure*}

We compute the map of the 3-$\sigma{}$ sensitivity limit using all the VLTI configurations displayed in Fig.~\ref{fig:visa}, assuming a target at declination $-35\,$deg. The limits are computed considering data sets consisting of respectively one pointing at an hour angle $\mathrm{HA}=-2\,$h, three pointings at hour angles $\mathrm{HA}=-2\,$h, $-1\,$h and $0\,$h, and five pointings at hour angles $\mathrm{HA}=-2\,$h, $-1\,$h, $0\,$h, $+1\,$h, and $+2\,$h. A closure phase accuracy of $0.25\deg$ is assumed (see Sect.~\ref{sec:accuracy}). We consider a maximum binary separation of $40\,$mas, which is the separation where AO-assisted spare aperture-masking imaging on 10-m class telescopes becomes more efficient than long-baseline interferometry.

Detailed results of the widest AT configuration (A1-K0-G1-I1) are displayed in Fig.~\ref{fig:det_map} for illustration. The figure shows that the detection completeness for a given sensitivity level increases drastically with the number of pointings. This is clearly illustrated by the decreasing number of ``blind spots'' (white zones) in the left-hand side plots. The sensitivity level is mostly flat for angular separations larger than 2\,mas, while the detection performance drops considerably within this inner working angle (IWA). The median sensitivity levels in the region 2--40 mas are respectively about $6\times 10^{-3}$, $4.5\times 10^{-3}$ and $4\times 10^{-3}$ for the three considered data sets. For more than five pointings, the median sensitivity level would continue to improve slightly, but the shape of the sensitivity curve would no longer significantly change.

The relative performances of the different configurations are illustrated in Fig.~\ref{fig:cont_sep} (left: sensitivity versus separation, right: completeness level versus contrast). All configurations provide flat performances for large separations, down to their respective IWA where the performances dramatically drop. The IWA are respectively $2\,$mas for the A1-K0-G1-I1 and U1-U2-U3-U4 configurations, $3\,$mas for the D0-H0-I1-G1 configuration, and $6\,$mas for the A1-B2-C1-D0 configuration. They correspond to the spatial resolution of the smallest baseline of the array. Given the similarity of the results for all configurations, we discuss them together in two different regimes: (i) the close-companion and (ii) the wide-companion regimes.

\subsection{Close-companion regime}

Close companions are defined here as companions with angular separations that are not fully resolved by \emph{at least} one baseline (that is $\vec{B}\cdot\vec{\Delta}/\lambda<1$). In this regime, the achievable contrast for a given completeness follows a power law of the angular separation $\ratio \propto \Delta^{-3}$, as predicted by Eq.~\ref{eq:close2}. The exact factor entering into this law depends on the array geometry but, as expected, the longest arrays provide the highest spatial resolution and are thus able to detect both the deepest and the closest binaries.

This well-known result was discussed by \citet{Lachaume03} in the context of partially resolved interferometric observations. We emphasize that our study additionally provides a quantitative estimation. As a typical example, we now detail the case of a faint companion with a contrast of $5\times10^{-3}$. We first consider the companion to be located at 2\,mas from the central star.  A closure-phase accuracy of $0.25\deg$ results in a detection efficiency of about 50\% using three pointings with the configuration A0-K0-G1-I1, according to Fig.~\ref{fig:det_map} (middle-right plot). However, if we now consider the companion to be located at $1\,$mas from the central star, the closure-phase accuracy should be $0.025\deg$ to reach the same efficiency. Since the angular separation is only marginally resolved by the interferometer, the lack of spatial resolution has to be compensated for by an increase in the accuracy on the signal (super-resolution effect).

\subsection{Wide-companion regime}

We note that those companions are \emph{wide} only in the interferometric sense, corresponding to separations larger than about 4\,mas for the typical $\sim100\,$m baselines available in modern interferometric facilities. 

In this regime, the detection efficiency becomes independent of the companion separation. Interestingly, all arrays have the same efficiency. In other words, as long as the companion is expected to be resolved by the interferometric baselines, the choice of array configuration does not matter. We conclude that there is no reason to favour a given VLTI configuration when looking for faint unknown companion with separations in the range $6-40\,$mas. More quantitatively, Fig.~\ref{fig:cont_sep} (right) displays the detection efficiency in this annular region for the four VLTI configurations versus the companion contrast, and for two observing scenarios (snapshot and long integration). The combination of five observations separated by one hour provides a detection efficiency higher than 95\% for companion contrasts of $10^{-2}$, assuming a realistic closure phase accuracy of $0.25\deg$. 

We note that the curve of completeness versus contrast become significantly sharper when accumulating observations. As shown by the solid lines in the right panel of Fig.~\ref{fig:cont_sep}, when accumulating five pointings, the efficiency drops from 80\% for a contrast of $5\times10^{-3}$ to less than 10\% for a contrast of $3\times10^{-3}$. The constraints provided by this dataset can thus be presented as a sensitivity limit and an inner working-angle, as for a classical imaging observation.

Quantitatively, when accumulating several pointings, these detection limits computed from the derivation of Sect.~\ref{sec:theory} are compatible with the blind-test analyses presented by \citet[Fig.~5]{Absil:2011} and the Monte-Carlo simulations of \citet[Fig.~4 and Eq.~2]{Lacour:2011}.

\subsection{Performances of standard configurations}
\label{sec:fake}

\begin{figure}
    \centering 
    \includegraphics[scale=0.58]{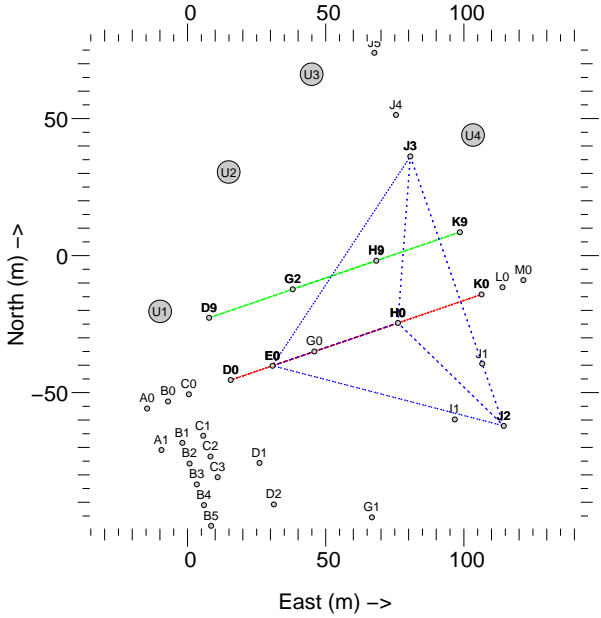}
    \caption{Fake VLTI interferometric configurations used in this paper: non-redundant linear D0-E0-H0-K0 (red), redundant linear D9-G2-H9-K9 (green, fake stations), and Y-shaped E0-J3-J2-H0 (blue).}
    \label{fig:fake}
\end{figure}

To add generality to the results presented in the previous section, we now study configurations that are not specifically linked to any existing interferometric array. We select the configurations presented in Fig.~\ref{fig:fake}, which is a non-redundant linear configuration, a fully redundant linear configuration and a Y-shaped configuration. All configurations have their longest baseline of the same size. The results are the following:

\begin{enumerate}
\item The linear non-redundant and Y-shaped configurations have similar detection limits as the currently offered (irregular) VLTI configurations for snapshot observations. Surprisingly, they also have similar performances when accumulating several pointings, while we may have expected that the Y-shape would unveil faster the remaining blind-spots.
\item For snapshot observations, the linear redundant configuration favours the \emph{dynamic range} with respect to the \emph{completeness}: it has a fainter detection limit for completeness levels below 50\%, but becomes significantly worse for higher completeness levels. When accumulating several pointings, both the highest completeness and largest dynamic range are reached, although the gain is never higher than 20\%.
\item Y-shaped arrays have smaller inner working angles than linear configurations of identical maximum baseline length, even when considering the accumulation of several pointings. The gain is almost a factor of two in terms of angular resolution.
\end{enumerate}

\begin{figure*}
    \centering 
    \includegraphics[scale=0.55]{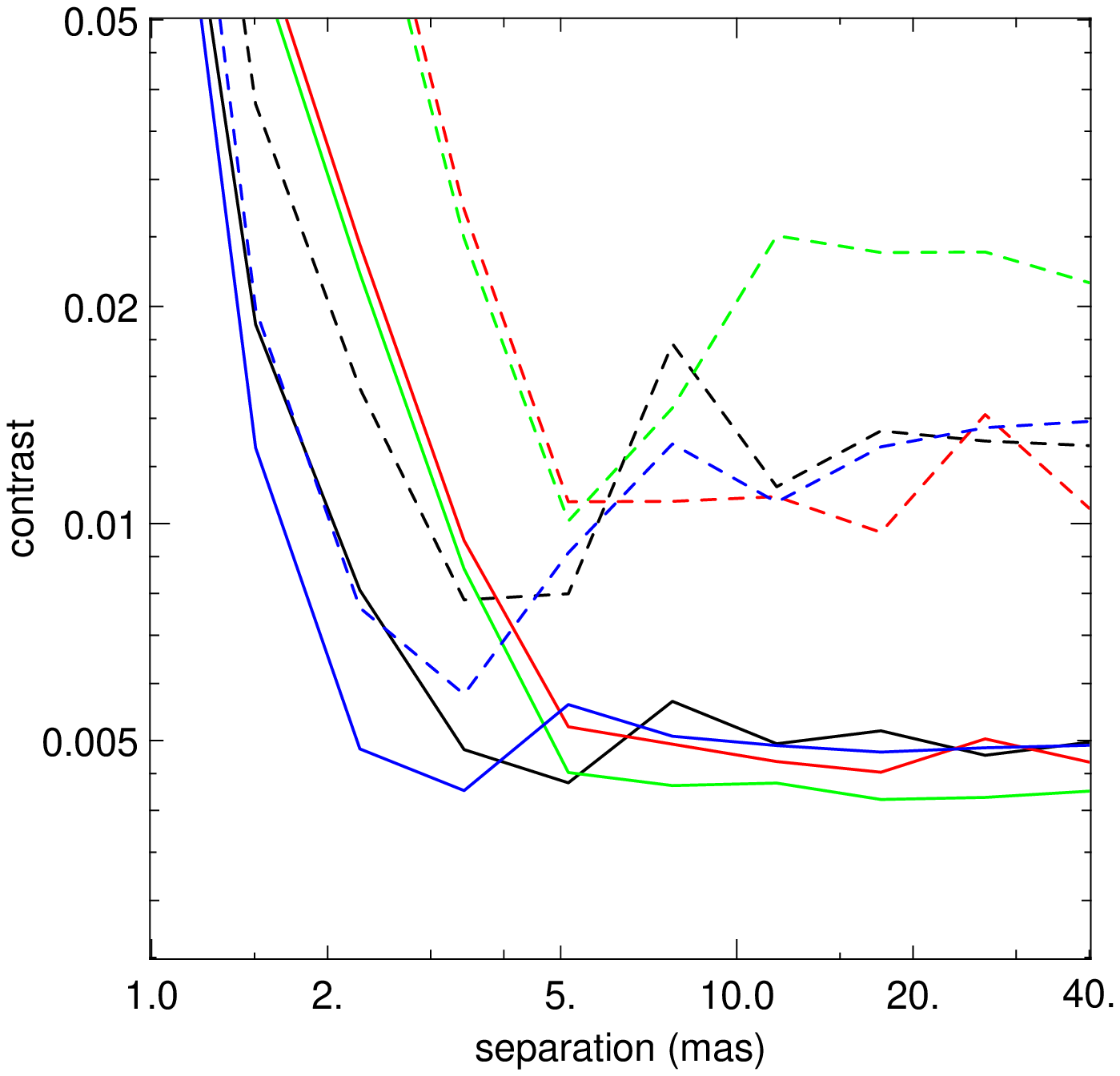}\hspace{2cm}
    \includegraphics[scale=0.55]{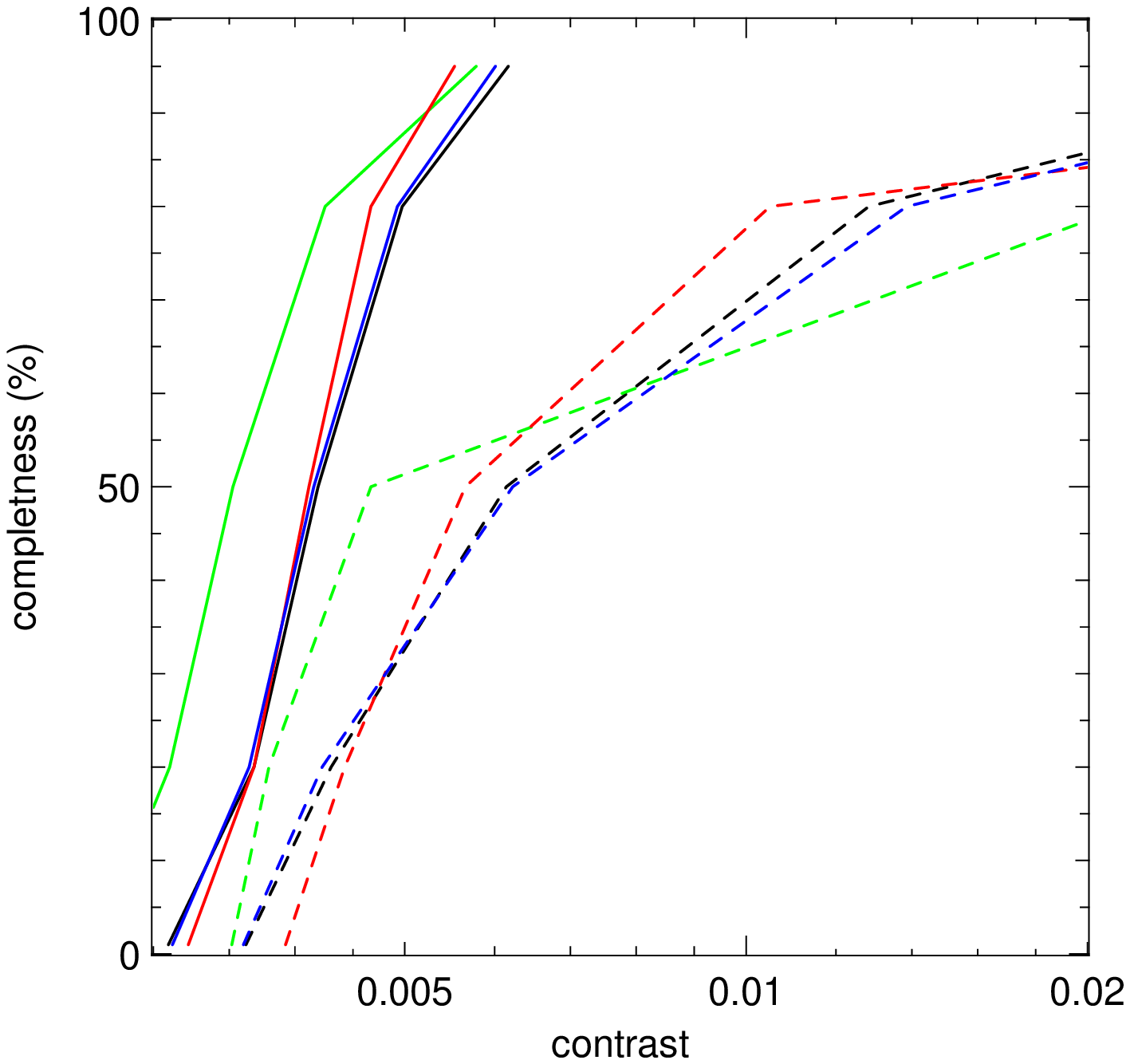}
    \caption{{\it Left:} 3-$\sigma{}$ sensitivity versus separation for a completeness level of 80\%. {\it Right:} Completeness in the separation range $6-40\,$mas versus contrast. Simulations are for a single snapshot pointing (dashed lines) and for five pointing separated by one hour (solid lines). Colours are for the four configurations displayed in Fig.~\ref{fig:fake}. The black curves are for the U1-U2-U3-U4 configuration displayed in Fig.~\ref{fig:visa}. The contrast axes can be scaled for any accuracy on the closure phase (here $\sigma=0.25\deg$).   }
    \label{fig:cont_sep_fake}
    \label{fig:comp_cont_fake}
\end{figure*}


\section{Closure phase accuracy and achievable dynamic range}
\label{sec:accuracy}

\subsection{Photon and piston noises}
We consider the theoretical photon noise limit for an observation of 1\,h on a star of sixth magnitude using one-metre class telescopes (e.g., the $1.8\,$m auxiliary telescopes of the VLTI). The choice of a sixth-magnitude star is driven by the current sensitivity limit of most interferometric instruments world-wide. A crude estimation of the photon noise is given by
\begin{equation}
\sigma_{phot} \approx \frac{360\deg}{\sqrt{N}} \; ,
\end{equation}
where $N$ is the total number of detected photons. If we consider that the 1\,h observing time should include the overheads and the observation of a calibration star, the effective integration time on target will be of the order of 20\,min. Assuming a realistic total transmission of 2\%, including both the instrumental and atmospheric contributions, the number of detected photons is $N\approx 10^8$ and the resulting photon noise is $\sigma_{phot} \approx 0.1\deg$. For a $1^\mathrm{mag}$ star, the resulting photon noise would be $\sigma_{phot} \approx 0.01\deg$.

In theory, closure phase is a robust observable against the telescope phase errors \citep{Monnier03}. However, in the context of non-zero exposure times and the presence of atmospheric turbulence, closure phase measurements are also affected by piston noise. A proper estimation of its amplitude is beyond the scope of this paper but it is still possible to provide a rough upper limit. Since piston noise is independent of the number of incident photons, it is expected to dominate the final statistical uncertainty for very bright stars. As an example, in the case of the PIONIER instrument at VLTI, the statistical uncertainty for bright stars is typically of the order of $0.25\deg$ to $2.5\deg$ for an integration time of 1\,min. This uncertainty depends on the atmospheric conditions as expected for piston noise. In decent atmospheric conditions, integrating over 20\,min allows the piston noise contribution to be reduced below $0.2\deg$, as it decreases with the square root of the integration time.

	\subsection{Calibration accuracy}

It is interesting to compare these fundamental limits to published accuracies that include the calibration of the instrumental closure phase (also called the transfer function):
\begin{description}
\item[VLTI/AMBER:] \citet{Absil:2010} reported calibration errors of between $0.20\deg$ and $0.37\deg$ depending on the night, using this three-telescope combiner in its medium spectral resolution mode ($R=1\,500$). With the low spectral resolution mode ($R=35$), typical calibration errors range from one to a few degrees \citep[see for instance][]{kraus:2009apr,le-bouquin:2009mar}.
\item[VLTI/PIONIER:] Typical calibration errors range from $0.25\deg$ to $1\deg$ \citep{LeBouquin:2011,Absil:2011} for this four-telescope combiner. Sequences with closure phases stable down to $0.1\deg$ have been recorded. Systematic discrepancies have been noted when calibration stars were separated by more than $10\deg$ on the sky.
\item[CHARA/MIRC:] The typical accuracy obtained with this four-telescope combiner is between $0.1$ and $0.2\deg$, which makes this instrument the most accurate of the currently available suite. Calibration uncertainties dominate the final accuracy at this level \citep{Zhao08,Zhao:2010,Zhao11}.
\end{description}

Altogether, a typical noise floor of $\sim0.25\deg$ seems to appear for the calibration of the closure phases in long baseline interferometry. Two results indicate that the major cause is probably longitudinal dispersion: (i) that the accuracy depends on the spectral resolution and (ii) the dependence on position of the calibration star on the sky. This is also the finding of \citet{Zhao11}, who proposed an elaborate calibration scheme for MIRC. Although this is clearly a very promising way of characterizing already known substellar companions, this method is probably not suited to surveying a large number of stars with a standard calibration procedure. Interestingly, $0.25\deg$ is also the noise floor reported by \citet{Lacour:2011} for the calibration of the closure phase of the spare aperture masking mode of NACO at VLT. This calibration noise floor of $0.25\deg$ theoretically does not prevent us from reaching very high dynamic ranges, by accumulating a large number of observations and/or baselines, as for instance in sparse aperture masking or spectrally dispersed observations. Care should however be taken to ensure that individual closure phase measurements are statistically independent. In particular, one should avoid repeating the same systematic errors in individual data sets, e.g., by choosing different calibrator stars, instrumental setups, etc., in order not to reach a true noise floor in the observations.

Concerning future instruments, the announced accuracy on the closure phases is $1\deg$ for the K-band four-telescope combiner GRAVITY \citep[document \mbox{VLT-SPE-ESO-15880-4853} and][]{Gillessen:2010} and from $1\deg$ to $5\deg$ for the L-band four-telescope combiner MATISSE \citep[Florentin Millour, private communication and][]{Lopez:2008}. Although these performances may be conservative, we conclude that the next generation of VLTI instruments is unlikely to break the $0.25\deg$ limit.

	\subsection{Discussion}

It appears realistic to reach a closure phase accuracy of $0.25\deg$ within less than one hour on one-metre class telescopes for stars of magnitude six and brighter. According to the results of Sect.~\ref{sec:array}, such performances allow a dynamic range of $5\times 10^{-3}$ ($\Delta \mathrm{mag}=5.75$) to be reached with 80\% completeness when five pointings are obtained with a four-telescope interferometer. The same performance would probably be reached within a snapshot using an interferometric instrument combining six telescopes or more at a time.

This result can be compared with the survey for stellar and sub-stellar companions on the ten-metre Keck and five-metre Palomar telescopes using aperture masking techniques in the K band \citep{Kraus08,Kraus:2011}. The achieved dynamic range is $\Delta K\approx5.5$ for separation as small as $25\,$mas (slightly worse for Palomar). A similar dynamic range and inner working angle have  been achieved within an ongoing survey of massive stars using aperture masking at VLT/NACO in the H band, e.g. $\Delta H\approx5$ down to $25\,$mas with this eight-metre diameter telescope (Hugues Sana, private communication). All close companions presented in these near-infrared surveys would have been detected by interferometry with an efficiency higher than 90\%, provided that they reside within the interferometric field-of-view. In addition, this efficiency would have been achieved down to about 2\,mas. At longer wavelengths, the aperture masking technique has a dynamic range of about $\Delta L\approx 7.5$ \citep{Hinkley:2011}, although the inner working angle in that case is only $70\,$mas. There is currently no L-band interferometric beam-combiner with closure phase capabilities to which these performances could be compared.

Several observing programs related to faint companion detection would benefit significantly from the capabilities of closure phase measurements on long-baseline interferometric instruments. An example is the search for low-mass (sub)-stellar companions around main-sequence stars residing in nearby young associations. Considering associations with ages between 10\,Myr and 200\,Myr, and a limiting magnitude $K=6$ for the instrument, one could survey stars up to about 15-20\,pc for stellar type M0V, 40\,pc for type G0V, and 120\,pc for type A0V. With an estimated median dynamic range of $\Delta K \simeq 6$, we computed the masses of the faintest companions that could be detected within a survey of nearby moving groups, using the (sub-)stellar cooling models of \citet{Baraffe98,Baraffe03}. The results are given in Table~\ref{tab:associations}, showing that the $13 M_{\rm Jup}$ ($=0.012 M_{\odot}$) limit between the brown dwarf and planetary regimes can be reached for young late-type dwarfs. In the case of A-type stars, one would be sensitive to companions in the range M3V-M7V depending on the age. For even younger stars, located in nearby star forming regions, closure phase measurements have the potential to reveal the formation of planetary-mass objects, as suggested by \citet{Kraus12}.

\begin{table}[t]
\caption{Sensitivity limits in terms of companion masses around young main-sequence stars.}
\centering
\begin{tabular}{c c c c}
\hline\hline
Age & A0V & G0V & M0V \\
\hline
10 Myr & $0.09 M_{\odot}$ & $0.017 M_{\odot}$ & $0.012 M_{\odot}$ \\
50 Myr & $0.22 M_{\odot}$ & $0.043 M_{\odot}$ & $0.013 M_{\odot}$ \\
200 Myr & $0.35 M_{\odot}$ & $0.08 M_{\odot}$ & $0.030 M_{\odot}$ \\
\hline
\end{tabular}
\label{tab:associations}
\end{table}

Another program is the determination of the binary fraction for massive stars. The interest is that despite the preponderance of multiple stars, the mechanism that produces multiple stars rather than single stars is still uncertain. The measurement of the mass distribution, and how it evolves with the mass of the primary, is an appropriate tool for disentangling between \emph{capture} and \emph{fragmentation} models. Stellar companions to B-type stars have been investigated using AO \citep[e.g. ][]{Roberts:2007}, although the stars observed typically have a large range of distances, limiting the statistical significance of the results. Radial velocity measurement of massive stars is challenging owing to the lack of suitable spectral lines and their intrinsic broadening. With the limiting magnitude $K=6$ presented in this paper, it is possible to observe interferometrically all the B stars within a distance of 75\,pc ($\sim$$100$ objects for the southern hemisphere), providing the first comprehensive study of massive binaries in the $0.25-5\,$AU separation range. 

Last but not least, one of the main selling argument for high-precision closure phases in optical interferometry has been the direct detection of hot extrasolar giant planets (EGP). Several hot EGP host stars are indeed bright enough to be observed with state-of-the-art interferometric instruments. For mature planetary systems ($>10$\,Myr), the expected contrast between the planet and the star is however generally too low ($<10^{-3}$) to be currently accessible with closure phase measurements \citep{Zhao08,Zhao11}. To routinely reach the hot EGP regime ($\Delta K \simeq 8 - 10$), a gain of two to four magnitudes is required in the dynamic range, which would translate into a noise floor of between $0.04\deg$ and $0.006\deg$ on the closure phase. Achieving such an accuracy would probably require a significant breakthrough in the instrumental domain.


\section{Conclusions}

In summary, optical interferometric surveys designed to detect faint companions have the following properties:
\begin{enumerate}
\item  The observable (closure phase) is robust against unstable atmospheric seeing conditions \citep{Monnier03}. Integrating over $20\,$min is sufficient to consistently reduce the photon and atmospheric noises below $0.25\deg$, which appears as a hard limit for the calibration of current instruments.
\item A single snapshot with four telescopes provides a 80\% detection efficiency at \mbox{$\Delta\mathrm{mag}=4.5$} as soon as the binary separation is fully resolved. The only requirement of the interferometric array is to use baselines as long as possible to improve the inner working angle, which is typically of the order of a few milli-arcseconds.
\item Accumulating more observations (several pointing and/or recombining more telescopes) allows a dynamic range \mbox{$\Delta\mathrm{mag}=6$} to be reached, which appears to be a realistic limit in respect to published performances. Going deeper would require us to break the current limit of $0.25\deg$ on the closure phase accuracy, or to massively increase the number of observations.
\item The achievable dynamic range scales linearly with the closure phase accuracy.
\end{enumerate}

In conclusion, interferometric closure phase surveys would be well-suited as filler programs for service-mode interferometric facilities, such as the VLTI. They can be considered as a useful complement to the AO-assisted imaging surveys currently carried out on ten-metre class telescopes. In particular, the search space of long-baseline interferometry bridges the gap between the wide companions found in direct imaging and the close companions detected by RV measurements. Moreover, interferometry could nicely complement RV studies in the particular cases where RV measurements are quite inappropriate. Young stars, for instance, are especially promising targets since their (sub)stellar companions are supposed to be relatively bright compared to their host stars.

\begin{acknowledgements} 
The authors thank the referee for his helpful comments. This work has made use of the Smithsonian/NASA Astrophysics Data System (ADS) and of the Centre de Donnees astronomiques de Strasbourg (CDS). All calculations and graphics were performed with the freeware \texttt{Yorick}\footnote{\texttt{http://yorick.sourceforge.net}}.
\end{acknowledgements}



\begin{thebibliography}{33}
\expandafter\ifx\csname natexlab\endcsname\relax\def\natexlab#1{#1}\fi

\bibitem[{{Absil} {et~al.}(2011){Absil}, {Le Bouquin}, {Berger}, {Lagrange},
  {Chauvin}, {Lazareff}, {Zins}, {Haguenauer}, {Jocou}, {Kern}, {Millan-Gabet},
  {Rochat}, \& {Traub}}]{Absil:2011}
{Absil}, O., {Le Bouquin}, J.-B., {Berger}, J.-P., {et~al.} 2011, \aap, 535,
  A68

\bibitem[{{Absil} {et~al.}(2010){Absil}, {Le Bouquin}, {Lebreton}, {Augereau},
  {Benisty}, {Chauvin}, {Hanot}, {M{\'e}rand}, \& {Montagnier}}]{Absil:2010}
{Absil}, O., {Le Bouquin}, J.-B., {Lebreton}, J., {et~al.} 2010, \aap, 520, L2+

\bibitem[{{Absil} \& {Mawet}(2010)}]{Absil10b}
{Absil}, O. \& {Mawet}, D. 2010, \aapr, 18, 317

\bibitem[{{Baraffe} {et~al.}(1998){Baraffe}, {Chabrier}, {Allard}, \&
  {Hauschildt}}]{Baraffe98}
{Baraffe}, I., {Chabrier}, G., {Allard}, F., \& {Hauschildt}, P.~H. 1998, \aap,
  337, 403

\bibitem[{{Baraffe} {et~al.}(2003){Baraffe}, {Chabrier}, {Barman}, {Allard}, \&
  {Hauschildt}}]{Baraffe03}
{Baraffe}, I., {Chabrier}, G., {Barman}, T.~S., {Allard}, F., \& {Hauschildt},
  P.~H. 2003, \aap, 402, 701

\bibitem[{{Bodenheimer} \& {Lin}(2002)}]{Bodenheimer02}
{Bodenheimer}, P. \& {Lin}, D.~N.~C. 2002, Annual Review of Earth and Planetary
  Sciences, 30, 113

\bibitem[{{Burrows}(2005)}]{Burrows05}
{Burrows}, A. 2005, \nat, 433, 261

\bibitem[{{Chelli} {et~al.}(2009){Chelli}, {Duvert}, {Malbet}, \&
  {Kern}}]{Chelli:2009}
{Chelli}, A., {Duvert}, G., {Malbet}, F., \& {Kern}, P. 2009, \aap, 498, 321

\bibitem[{{Duvert} {et~al.}(2010){Duvert}, {Chelli}, {Malbet}, \&
  {Kern}}]{Duvert:2010}
{Duvert}, G., {Chelli}, A., {Malbet}, F., \& {Kern}, P. 2010, \aap, 509, A66+

\bibitem[{{Gillessen} {et~al.}(2010){Gillessen}, {Eisenhauer}, {Perrin},
  {Brandner}, {Straubmeier}, {Perraut}, {Amorim}, {Sch{\"o}ller},
  {Araujo-Hauck}, {Bartko}, {Baumeister}, {Berger}, {Carvas}, {Cassaing},
  {Chapron}, {Choquet}, {Clenet}, {Collin}, {Eckart}, {Fedou}, {Fischer},
  {Gendron}, {Genzel}, {Gitton}, {Gonte}, {Gr{\"a}ter}, {Haguenauer}, {Haug},
  {Haubois}, {Henning}, {Hippler}, {Hofmann}, {Jocou}, {Kellner}, {Kervella},
  {Klein}, {Kudryavtseva}, {Lacour}, {Lapeyrere}, {Laun}, {Lena}, {Lenzen},
  {Lima}, {Moratschke}, {Moch}, {Moulin}, {Naranjo}, {Neumann}, {Nolot},
  {Paumard}, {Pfuhl}, {Rabien}, {Ramos}, {Rees}, {Rohloff}, {Rouan}, {Rousset},
  {Sevin}, {Thiel}, {Wagner}, {Wiest}, {Yazici}, \& {Ziegler}}]{Gillessen:2010}
{Gillessen}, S., {Eisenhauer}, F., {Perrin}, G., {et~al.} 2010, in Presented at
  the Society of Photo-Optical Instrumentation Engineers (SPIE) Conference,
  Vol. 7734, Society of Photo-Optical Instrumentation Engineers (SPIE)
  Conference Series

\bibitem[{{Hinkley} {et~al.}(2011){Hinkley}, {Carpenter}, {Ireland}, \&
  {Kraus}}]{Hinkley:2011}
{Hinkley}, S., {Carpenter}, J.~M., {Ireland}, M.~J., \& {Kraus}, A.~L. 2011,
  \apjl, 730, L21+

\bibitem[{{Kraus} \& {Ireland}(2012)}]{Kraus12}
{Kraus}, A.~L. \& {Ireland}, M.~J. 2012, \apj, 745, 5

\bibitem[{{Kraus} {et~al.}(2011){Kraus}, {Ireland}, {Martinache}, \&
  {Hillenbrand}}]{Kraus:2011}
{Kraus}, A.~L., {Ireland}, M.~J., {Martinache}, F., \& {Hillenbrand}, L.~A.
  2011, \apj, 731, 8

\bibitem[{{Kraus} {et~al.}(2008){Kraus}, {Ireland}, {Martinache}, \&
  {Lloyd}}]{Kraus08}
{Kraus}, A.~L., {Ireland}, M.~J., {Martinache}, F., \& {Lloyd}, J.~P. 2008,
  \apj, 679, 762

\bibitem[{{Kraus} {et~al.}(2009){Kraus}, {Weigelt}, {Balega}, {Docobo},
  {Hofmann}, {Preibisch}, {Schertl}, {Tamazian}, {Driebe}, {Ohnaka}, {Petrov},
  {Sch{\"o}ller}, \& {Smith}}]{kraus:2009apr}
{Kraus}, S., {Weigelt}, G., {Balega}, Y.~Y., {et~al.} 2009, \aap, 497, 195

\bibitem[{{Lachaume}(2003)}]{Lachaume03}
{Lachaume}, R. 2003, \aap, 400, 795

\bibitem[{{Lacour} {et~al.}(2008){Lacour}, {Meimon}, {Thi{\'e}baut}, {Perrin},
  {Verhoelst}, {Pedretti}, {Schuller}, {Mugnier}, {Monnier}, {Berger},
  {Haubois}, {Poncelet}, {Le Besnerais}, {Eriksson}, {Millan-Gabet}, {Ragland},
  {Lacasse}, \& {Traub}}]{Lacour08}
{Lacour}, S., {Meimon}, S., {Thi{\'e}baut}, E., {et~al.} 2008, \aap, 485, 561

\bibitem[{{Lacour} {et~al.}(2011){Lacour}, {Tuthill}, {Amico}, {Ireland},
  {Ehrenreich}, {Huelamo}, \& {Lagrange}}]{Lacour:2011}
{Lacour}, S., {Tuthill}, P., {Amico}, P., {et~al.} 2011, \aap, 532, A72+

\bibitem[{{Le Bouquin} {et~al.}(2011){Le Bouquin}, {Berger}, {Lazareff},
  {Zins}, {Haguenauer}, {Jocou}, {Kern}, {Millan-Gabet}, {Traub}, {Absil},
  {Augereau}, {Benisty}, {Blind}, {Bonfils}, {Bourget}, {Delboulbe},
  {Feautrier}, {Germain}, {Gitton}, {Gillier}, {Kiekebusch}, {Kluska},
  {Knudstrup}, {Labeye}, {Lizon}, {Monin}, {Magnard}, {Malbet}, {Maurel},
  {M{\'e}nard}, {Micallef}, {Michaud}, {Montagnier}, {Morel}, {Moulin},
  {Perraut}, {Popovic}, {Rabou}, {Rochat}, {Rojas}, {Roussel}, {Roux},
  {Stadler}, {Stefl}, {Tatulli}, \& {Ventura}}]{LeBouquin:2011}
{Le Bouquin}, J.-B., {Berger}, J.-P., {Lazareff}, B., {et~al.} 2011, \aap, 535,
  A67

\bibitem[{{Le Bouquin} {et~al.}(2009){Le Bouquin}, {Lacour}, {Renard},
  {Thi{\'e}baut}, {Merand}, \& {Verhoelst}}]{le-bouquin:2009mar}
{Le Bouquin}, J.-B., {Lacour}, S., {Renard}, S., {et~al.} 2009, \aap, 496, L1

\bibitem[{{Lopez} {et~al.}(2008){Lopez}, {Antonelli}, {Wolf}, {Lagarde},
  {Jaffe}, {Navarro}, {Graser}, {Petrov}, {Weigelt}, {Bresson}, {Hofmann},
  {Beckman}, {Henning}, {Laun}, {Leinert}, {Kraus}, {Robbe-Dubois}, {Vakili},
  {Richichi}, {Abraham}, {Augereau}, {Behrend}, {Berio}, {Berruyer},
  {Chesneau}, {Clausse}, {Connot}, {Demyk}, {Danchi}, {Dugu{\'e}}, {Finger},
  {Flament}, {Glazenborg}, {Hannenburg}, {Heininger}, {Hugues}, {Hron},
  {Jankov}, {Kerschbaum}, {Kroes}, {Linz}, {Lizon}, {Mathias}, {Mathar},
  {Matter}, {Menut}, {Meisenheimer}, {Millour}, {Nardetto}, {Neumann},
  {Nussbaum}, {Niedzielski}, {Mosoni}, {Olofsson}, {Rabbia}, {Ratzka}, {Rigal},
  {Roussel}, {Schertl}, {Schmider}, {Stecklum}, {Thiebaut}, {Vannier}, {Valat},
  {Wagner}, \& {Waters}}]{Lopez:2008}
{Lopez}, B., {Antonelli}, P., {Wolf}, S., {et~al.} 2008, in Presented at the
  Society of Photo-Optical Instrumentation Engineers (SPIE) Conference, Vol.
  7013, Society of Photo-Optical Instrumentation Engineers (SPIE) Conference
  Series

\bibitem[{{Monnier}(2003)}]{Monnier03}
{Monnier}, J.~D. 2003, Reports on Progress in Physics, 66, 789

\bibitem[{{Monnier} {et~al.}(2006){Monnier}, {Pedretti}, {Thureau}, {Berger},
  {Millan-Gabet}, {ten Brummelaar}, {McAlister}, {Sturmann}, {Sturmann},
  {Muirhead}, {Tannirkulam}, {Webster}, \& {Zhao}}]{Monnier:2006}
{Monnier}, J.~D., {Pedretti}, E., {Thureau}, N., {et~al.} 2006, in Society of
  Photo-Optical Instrumentation Engineers (SPIE) Conference Series, Vol. 6268,
  Society of Photo-Optical Instrumentation Engineers (SPIE) Conference Series

\bibitem[{{Oppenheimer} \& {Hinkley}(2009)}]{Oppenheimer09}
{Oppenheimer}, B.~R. \& {Hinkley}, S. 2009, \araa, 47, 253

\bibitem[{{Perryman}(2000)}]{Perryman00}
{Perryman}, M.~A.~C. 2000, Reports on Progress in Physics, 63, 1209

\bibitem[{{Renard} {et~al.}(2008){Renard}, {Absil}, {Berger}, {Bonfils},
  {Forveille}, \& {Malbet}}]{Renard08}
{Renard}, S., {Absil}, O., {Berger}, J.-P., {et~al.} 2008, in Proc. SPIE, Vol.
  7013, Optical and Infrared Interferometry, ed. M.~{Sch\"oller}, W.~{Danchi},
  \& F.~{Delplancke}

\bibitem[{{Roberts} {et~al.}(2007){Roberts}, {Turner}, \& {ten
  Brummelaar}}]{Roberts:2007}
{Roberts}, Jr., L.~C., {Turner}, N.~H., \& {ten Brummelaar}, T.~A. 2007, \aj,
  133, 545

\bibitem[{{Schneider} {et~al.}(2011){Schneider}, {Dedieu}, {Le Sidaner},
  {Savalle}, \& {Zolotukhin}}]{Schneider11}
{Schneider}, J., {Dedieu}, C., {Le Sidaner}, P., {Savalle}, R., \&
  {Zolotukhin}, I. 2011, \aap, 532, A79

\bibitem[{{Udry} \& {Santos}(2007)}]{Udry07}
{Udry}, S. \& {Santos}, N.~C. 2007, \araa, 45, 397

\bibitem[{{Vannier} {et~al.}(2006){Vannier}, {Petrov}, {Lopez}, \&
  {Millour}}]{Vannier06}
{Vannier}, M., {Petrov}, R.~G., {Lopez}, B., \& {Millour}, F. 2006, \mnras,
  367, 825

\bibitem[{{Zhao} {et~al.}(2011){Zhao}, {Monnier}, {Che}, {Pedretti}, {Thureau},
  {Schaefer}, {Ten Brummelaar}, {M{\'e}rand}, {Ridgway}, {McAlister}, {Turner},
  {Sturmann}, {Sturmann}, {Goldfinger}, \& {Farrington}}]{Zhao11}
{Zhao}, M., {Monnier}, J.~D., {Che}, X., {et~al.} 2011, \pasp, 123, 964

\bibitem[{{Zhao} {et~al.}(2010){Zhao}, {Monnier}, {Che}, {Ten Brummelaar},
  {Pedretti}, \& {Thureau}}]{Zhao:2010}
{Zhao}, M., {Monnier}, J.~D., {Che}, X., {et~al.} 2010, in Society of
  Photo-Optical Instrumentation Engineers (SPIE) Conference Series, Vol. 7734,
  Society of Photo-Optical Instrumentation Engineers (SPIE) Conference Series

\bibitem[{{Zhao} {et~al.}(2008){Zhao}, {Monnier}, {ten Brummelaar}, {Pedretti},
  \& {Thureau}}]{Zhao08}
{Zhao}, M., {Monnier}, J.~D., {ten Brummelaar}, T., {Pedretti}, E., \&
  {Thureau}, N.~D. 2008, in Proc. SPIE, Vol. 7013, Optical and Infrared
  Interferometry, ed. M.~{Sch\"oller}, W.~{Danchi}, \& F.~{Delplancke}

\end{thebibliography}


\end{document}